\documentclass[12pt]{article}

\usepackage{graphicx}
\usepackage{dcolumn}
\usepackage{bm}
\usepackage{amsmath}
\usepackage{color}
\usepackage{amssymb}
\usepackage{url}

\begin{document}

   \renewcommand{\topfraction}{.9}
   \renewcommand{\bottomfraction}{.9}
   \renewcommand{\textfraction}{.1}

\title{Slow, Continuous Beams of Large Gas Phase Molecules}

\author{David Patterson and John M. Doyle}

\date{\today}

\begin{abstract}

Cold, continuous, high flux beams of benzonitrile, fluorobenzine, and anisole have been created. Buffer-gas cooling with a cryogenic gas provides the cooling and slow forward beam velocities.  The beam of benzonitrile was measured to have a forward velocity peaked at 67 $\pm 5$  m s$^{-1}$, and a continuous flux of $10^{15}$ molecules s$^{-1}$. These beams provide a continuous source for high resolution spectroscopy, and provide an attractive starting point for further spatial manipulation of such molecules, including eventual trapping.

\end{abstract}

\maketitle

\section{\label{sec:level1}Introduction}

Cold molecules are rich quantum structures with a variety of potential applications, such as fundamental physics tests~\cite{astromethanol} and quantum information processing~\cite{ikenmr,molensemblepreprint}. 
Cooling and quantum control of polyatomics lags far behind that of diatomics, which are now routinely manipulated at the single quantum state level~\cite{Jincolddipolar}. Larger polyatomic polar molecules have some advantages, such as larger dipole moments and richer internal modes (important for quantum computing and simulation), but control of the many degrees of freedom available in larger molecules remains challenging. 
  Demonstrated sources of larger ($>$ 2 atoms) cold molecules include AG-decelerated samples from supersonic jets and (internally warm) cold molecules filtered from warm sources \cite{rempefilter}.  Molecules as large as H$_2$CO have previously been extracted from cryogenic buffer gas sources\cite{buurenrempe2008}. Rempe et. al. have recently demonstrated Sysiphus-type cooling of a trapped sample of CH$_3$F molecules, producing the coldest sample of trapped polyatomic molecules to date. Here we present a novel, continuous beam of cold molecules based on the buffer gas cooling of a room temperature source to a few degrees K.  The modest forward velocity of the source ($\approx$ 70 m s$^{-1}$) makes it an attractive starting point for further manipulation and trapping.\cite{Junyeoverview}\cite{Meijertunable}\cite{MeijerReview} The molecules from this source could be used without further manipulation for precision spectroscopy, or slowed and trapped for further cooling\cite{hsinicaf}.

 \section{Experimental}

\begin{figure}
\includegraphics[width = 150mm]{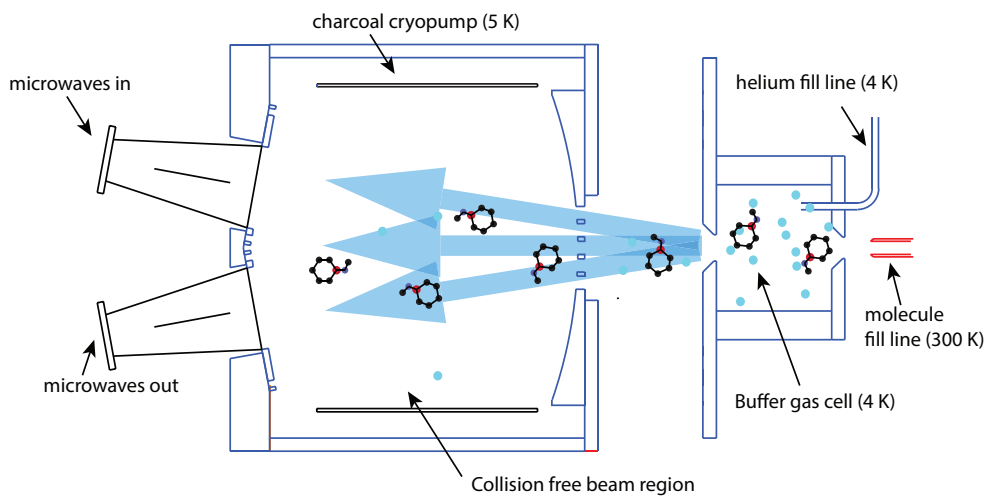}
\caption{The apparatus to produce a cold, continuous, slow beam of larger molecules.  Molecules are introduced to a cold cell via a warm injection tube.  The molecules cool in the cell via collisions with cold helium buffer gas, and a fraction escape through the exit aperture.  These molecules spray towards a second aperture, which leads into a separately cryopumped chamber where they are interrogated via Fourier transform microwave spectroscopy.  The second aperture is covered by a course mesh, which is largely transparent to the molecular beam but reflects microwaves. The mean free time between helium collisions is measured at 15 $\mu$s in the cell, and estimated to be more than 1000 $\mu$s in the beam chamber.}
\label{apparatusfig}
\end{figure}

The essential approach to our new beam is described in reference \cite{ourslowbeam}, but with significant modifications to the molecular injection, adapting it to stable, polyatomic molecules.
A diagram of our apparatus is shown in figure~\ref{apparatusfig}.  The beam cell used here is a modification of that employed in our earlier closed-cell work\cite{mefirstFTMW}, which produced, inside a cell and mixed with cold buffer gas, continuous gas phase samples of benzonitrile, acetone, 1-2 propanediol, fluorobenzene, and anisole.  As in that work, here hot molecules are injected by spraying rotationally and vibrationally warm molecules from a hot pipe towards a 1 cm aperture in a cold (4.7 K) cell.  Molecules are cooled through collisions with the cold helium gas. In contrast to other geometries in which hot molecules are introduced via a capillary connected directly to a buffer gas cell\cite{patterson2007,buurenrempe2008}, this non-contact approach minimizes heat loads and allows for a much larger injection pipe, which is critical for larger, less volatile species.  The cell contains cold helium gas with a low enough density such that warm molecules can penetrate through the gas curtain exiting the cold aperture, but a high enough density such that the hot molecules are thermalized via collisions with helium once they are inside before exiting the cell or hitting and sticking to the copper walls of the cell. This ``sweet spot'' density is  $n_{He} \approx 4 \pm 2 \times 10^{15}$ cm$^{-3}$.  A fraction of the resulting cold molecules, along with cold helium gas, exits the cell via a new second aperture on the opposite side of the cell, into a vacuum region. Once in the vacuum region, the density of helium (and molecules) drops rapidly as the gas mixture moves away from the cell, realizing a beam.  This beam passes into a separately cryopumped ``spectroscopy chamber,'' a collision free beam region.

The molecules are detected in the spectroscopy chamber and their forward velocity is measured via Fourier transform microwave spectroscopy.  This is realized in a retro-reflecting microwave beam geometry.  The microwave field, which has a direction of propagation close to parallel to the molecular beam axis, is emitted by a standard gain microwave horn (WR-62), and is reflected and refocused by a spherical mirror onto a second microwave horn.  This second horn is used to detect the free induction decay signal from the molecules after they are polarized via a strong, chirped microwave pulse.  As in standard COBRA-type FTMW spectrometers, each rotational transition of the molecules exhibits a ``doppler doublet'' type structure, resulting from interactions with the co-propagating and counter-propagating components of the applied microwave field with the forward moving molecules.  The velocity distribution of the beam is determined from the doppler-broadened structure of each lobe of this doppler-doublet. We call this the ``rabbit geometry''.

 \section{Results}
\begin{figure}
\includegraphics[width = 150mm]{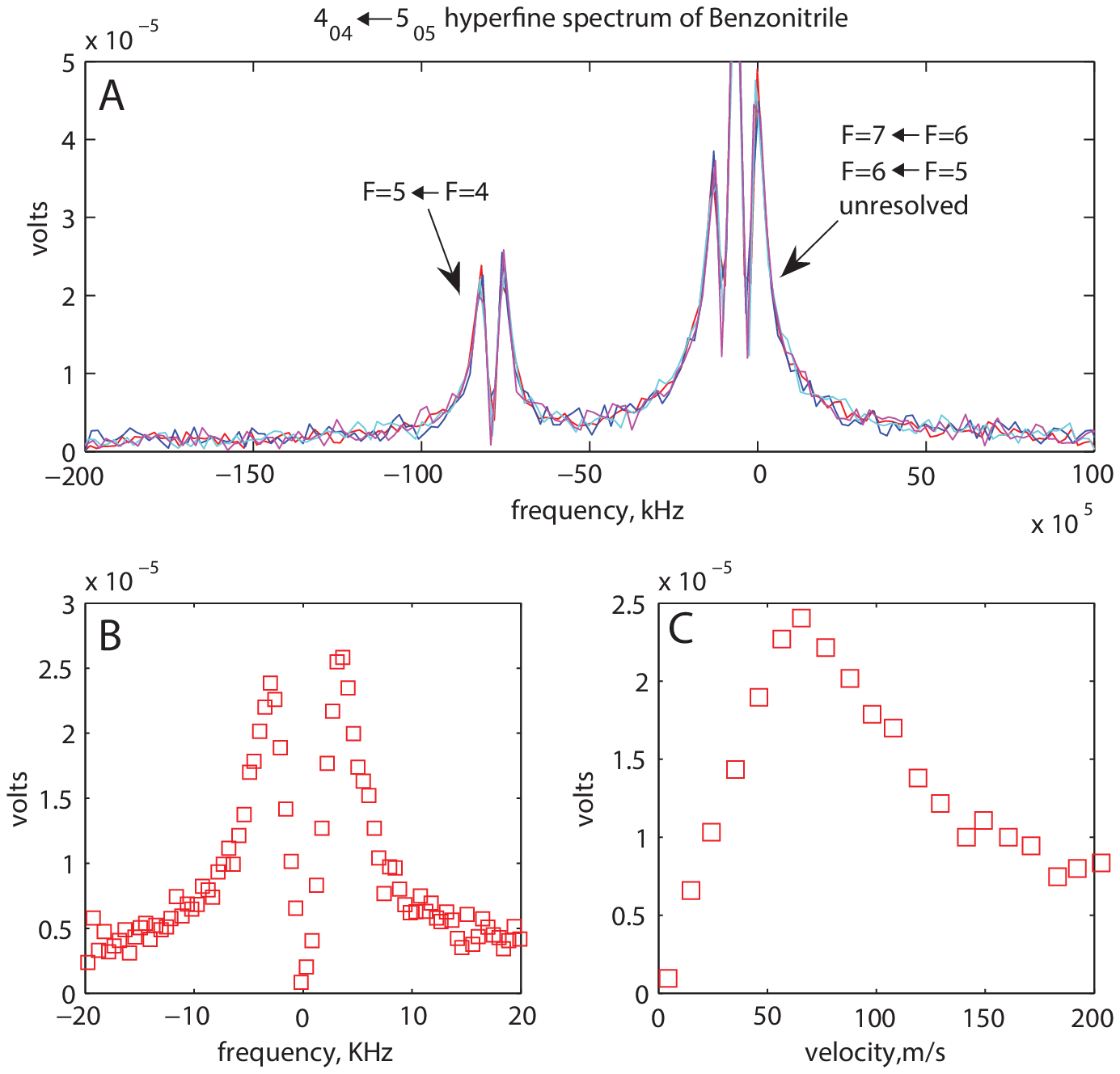}
\caption{The microwave spectrum of a beam of cold benzonitrile. A: The partial hyperfine manifold of the $4_{04} \leftarrow 5_{05}$ transition at 13437.4 MHz.  Each hyperfine line is split into a doppler doublet, as is seen in COBRA-configured cavity enhanced microwave spectrometers with a supersonic beam source.  B: An enlarged view of the resolved $F=5\rightarrow F=4$ double. C: the velocity distribution for the beam, as measured from the red sideband of the $F=5 \leftarrow F=4$ doublet.  The beam has a peak velocty of 68 m s$^{-1}$, and an estimated flux of $10^{15}$ molecules s${-1}$.}
\label{datafig}
\end{figure}

Figure~\ref{datafig} shows the spectrum from a cold beam of benzonitrile, C$_6$H$_5$CN.  In figure~\ref{datafig}a, the $3_{03} \leftarrow 4_{04}$ hyperfine manifold is shown; while the strong peaks corresponding to the $F=7 \leftarrow F=6$ and $F=6 \leftarrow F=5$ are only partially resolved, the $F=5 \leftarrow F=4$  transition at 13437.300 MHz is resolved cleanly.  Figure~\ref{datafig}c shows the velocity distribution derived from the red sideband of the $F=5 \leftarrow F=4$ transition.

 The forward velocity distribution is peaked at $v_p = 67 \pm 5$ m $s^{-1}$,  corresponding to a forward kinetic energy of 28 K, and has translational temperature of 10 kelvin in the moving frame, although it should be emphasized that this velocity distribution is rather poorly described by a Maxwell-Boltzman distribution. The forward velocity spread of the beam is $\approx \pm$ 40m $s^{-1}$.  It is natural to compare $v_p$ to a more commonly used source of cold molecules, a seeded pulsed supersonic jet.  Such jets typically have a forward velocity on the order of 400 m $s^{-1}$ or greater, corresponding to a kinetic energy of 1000 K or more, combined with a moving frame energy spread (i.e. translational temperature) of 2 K or less. Depending on the experiment one wants to accomplish with a cold beam of molecules, this difference in parameters can be crucial. We consider one application just below, deceleration and trapping of molecules.

~

\noindent
{\emph{Discussion}}

A powerful tool for slowing and trapping diatomic molecules is the electric field interaction with the molecular dipole. Manipulation of larger molecules via electric fields is complicated by the fact that, unlike smaller molecules, at high fields such molecules exhibit only high field seeking states. In addition, rotationally excited molecules in changing electric fields undergo numerous level crossings and anti-crossings, making control extremely challenging.  In particular, this means that conventional low-field seeking stark decelerators and electrostatic traps have very low effective depths. Considerable progress has been made in recent years in developing both alternating gradient decelerators\cite{BethlemAG,JochenAG} and microwave decelerators\cite{SchnellDecelerate} adapted for high field seeking states; for an excellent review of the field see reference \cite{manipulateoverview}.  Two possible types of decelerator one can consider are a high field seeking decelerator, which can decelerate ground state (high field seeking) molecules with a single, strong electric field, and a staged decelerator with many weak, switched fields.  In both cases molecules lose kinetic energy as they run ``uphill'' in the switched potential.
For the single stage decelerator effective for ground state molecules, it is natural to compare the kinetic energy of a molecule from a beam source to the available ``stopping'' energy from an electric field, $E\cdot D$, where $D$ is the molecular dipole.  For a realistic applied field of 100 kV/cm interacting with
 benzonitrile ($D$ = 4.5 Debye), $E\cdot D$ corresponds to 10.9 K, which in turn corresponds to a velocity of 42 m s$^{-1}$.  The forward velocity distribution shown in figure~\ref{datafig}c therefore contains a significant fraction of molecules that could be decelerated to zero via a single switched electric field. Alternatively, a switched, ``travelling wave'' low field seeking decelerator with modest fields $E_{max}$ and depth $D$, and number of stages $N$ (for benzonitrile, $E_{max} = 3 $ kV/cm and $D = $50 mK, $N \approx 100$) seems to be possible \cite{hoekstratw}.  Such a decelerator has the advantage of stability even at zero velocity (i.e. trapping), but requires more elaborate engineering.  Both designs would be impractical without the low lab frame kinetic energy of the buffer gas cooled beam.


The general nature of our source and the tunability of microwave detection make cooling other species straightforward.  Beams of anisole, aminobenzonitrile, and fluorobenzene (aromatics) with similar characteristics to the one showed here were also produced in our apparatus.  Surprisingly, our attempts to produce  beams of acetone and 1-2 propanediol did not result in a detectable signal, even though high density samples of these molecules were observed in the cell.  We hypothesize that this relates to an at-present not understood clustering mechanism which could sweep up monomers of these species before they migrate out of the cell and into the beam. This observation remains for further investigation.

\section{Conclusion}

We have created slow beams of several molecules and performed chirped pulse microwave spectroscopy in a novel cryogenic beam geometry. The typical average forward velocity of the beam is 70 m/s with a substantial number of molecules with velocity below 40 m/s. At such a velocity, transit time broadening is substantially reduced compared to supersonic beams, thus making this approach potentially useful for higher resolution beam spectroscopy.  The low kinetic energy of these beams makes them an attractive starting point for further spatial manipulation.

\bibliographystyle{amsplain}
\bibliography{beambib}

\end{document}